## I. Introduction

The exposure of EMF is increasing compared to previous generations due to 5G has been deployed in many countries for fast and reliable data transfer. For low latency 5G uses Ultra-Reliable and Low-Latency Communications. It uses enhanced Mobile Broad Band. From 1G to 5G, the penetration rates increased dramatically to provide high data rates. The access points of 5G are increasing nowadays since the usage of wireless devices is increased. It will raise the level of exposure and EMF radiation and it impacts on human body. 5G provides many connections using massive Machine Type Communications. Several researchers are analysing the effects of EMF on 5G networks. International standards such as WHO, ICNIRP, ITU, and FCC provided the guidelines to protect humans from EMF. Increasing the UE by humans will raise questions about their health issues. And long-term use of mobiles may create several health impacts, such as cancer, sperm fertility, gland tumour, cataract, and many skin diseases.

The biggest question raised about 5G is that higher frequencies will affect health and the environment. The skin is a thick, big, delicate, and easily damaged organ. Since human skin can block the higher frequencies of sunlight, the risk of skin cancer will increase by 5G. Since the skin is the body's outermost layer and most significant organ, it is a primary concern for the health effects of EMF—various symptoms or diseases caused by EMF or non-ionizing radiations. The most frequent disease is skin pigmentation. The widely used frequency bandwidth of 5G communication devices is 28 GHz. It is a substantial public concern about the lack of information about human safety and environment [7]. Since many transmitters are operating in 5G, they need to deploy more BS due to the small cells' employment. The Base Stations are located closer to users, cover smaller geographical areas, and lead to high exposure to EMF. The purpose of using multiple antennas is to increase the antenna's gain. It results in penetrating EMF into the human body potentially [14]. This paper discussed about the penetration of RF-EMF of 5G wireless network in human skin. The metric to measure the penetration level 5G in human skin is Specific Absorption Rate. It shows the energy quantity that the body observes from EMF.

## II. Literature Review

Kyurikim et al. observed that the melanin and ROS are produced by 5G exposure. But, need some long-term studies about melanin synthesis such as αMSH, and PMBT of shorter wavelengths [1]. Kyurikim et al. investigated skin pigmentation in vitro with EMF effects of LTE and 5G (28 GHz) [2]. They examined about the impact of B16F10 murine melanoma cell line. Mehdizadeh et al. discussed the of 5G technology safety controversies. The penetration depth of 5G EMF in human tissue lies in a frequency of 10GHz. High-frequency RF-EMF be able to penetrate the human skin and damage the living skin cells severely [3]. Yakymenko et al. conducted studies of about 100 peer-review regarding the low-intensity RF radiation oxidative effects, 93 studies have given the positive results that RF induced oxidative impacts in biological systems [4]. Mortazani et al. reported no link between RF-EMF exposure and brain cancer [5]. Haghani M et al. presented a relationship of nonlinear J-shaped dose-response for non-ionizing RF-EMF carcinogenesis. It reduces the risk fear associated with RF-EMF exposures at deficient levels. It shows the hermetic effects at low levels and the possibilities of cancer and irreversible damage at high levels [6]. The biological influences on human skin have been increased expression of mitogenic signal transduction genes, oxidative stress, and increased DNA synthesis. Mobile users have diagnosed the symptoms are fatigue, headache, tingling, itching, dizziness, and warm sensation [13]. Seungmo Kim et al. calculates the human EMF exposure and the penetration level into human skin in 5G at 28 GHz [14]. The scheduling tool evaluates exposure and optimization in WiFi or LTE and AP femtocell – provided indoor wireless networks [17]. Captain Jerry G. Flynn wrote in Hidden Dangers 5G, released on November 13, 2019, that 5G will result in two billion deaths.

### III. Electro Magnetic Field

Electromagnetic radiation (EMR) is used by doctors in a variety of ways. Pulsed EMR are becoming more popular in orthopaedics and physical therapy. For cancer treatment and diagnosis high-frequency ionizing EMR is used. To treat some skin diseases like psoriasis, use non-ionizing UV radiation. Visible light makes up a small portion of the electromagnetic spectrum Figure 1, with most of the other portions being unseen. Ionizing radiation is a higher energy part of the UV region of the spectrum. It can be able to remove the electrons from atoms and breaks chemical bonds, resulting in damage or killing the cells, and can change the structure of DNA. Non-ionizing radiation can cause skin damage, hyperpigmentation, cataracts, and photoaging when extreme heat is released. The dangerous ultraviolet (UV) radiation is found in the UVA and UVB frequency bands. Recent research indicates that UVA playing a bigger role in the development of skin cancer than was previously thought and the immune system and other organs may be impacted. This contrasts with the conventional theory that UVB causes skin cancer and direct DNA damage that leads to melanoma. Other portions of the spectrum, in particular blue light at 440 nanometres, are employed due to its biological effects on the skin to treat hyperbilirubinemia in jaundiced infants.

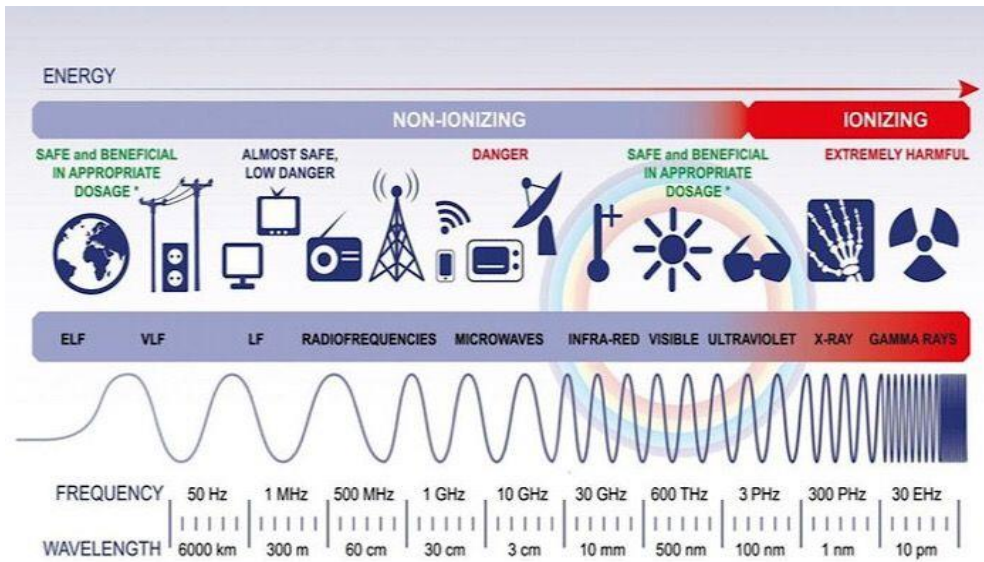

*Figure 1Ionizing and Non-Ionizing Radiation of EMF (HELIOS3, 2023)*

By promoting the liver's production of di-hydroxyvitamin D, this is accomplished. Acute bilirubin encephalopathy and kernicterus, two neurologic diseases that are profoundly debilitating, can develop as a result of the syndrome if bilirubin concentrations are not treated. Blue light is also known to disrupt sleep by preventing the pineal gland from producing melatonin, a powerful antioxidant and free radical scavenger that is created when you sleep in the dark.

5G Spectrum Bands

Expanding into the new spectrum, it provides more network capacity. 5G operates on three spectrum bands. First, low-band waves range from 600MHz to 900MHz and can travel long distances. The use cases of low bands are the IoT industry, logistics, low-frequency monitoring in smart cities, and monitoring in smart agriculture. Second, mid-band waves from 2.5GHz to 4.2GHz, and the use cases are media and entertainment, the health industry, public safety surveillance, autonomous drones, transportation in smart cities, and the self-driving vehicle in smart agriculture. The third, high bands, also called mm waves, are shorter waves, so they cannot penetrate trees, walls, buildings, etc., and the use cases are manufacturing, media and entertainment, automobile, and retail. 5G belongs to fixed wireless access. The frequency spectrum bands of 5G are depicts in Table 1. 5G supports two frequency ranges of operations.

Table 1 5G Frequency Spectrum Bands

| Bands | Freq. Range | Extension | Cells | Use | Comments |
|---|---|---|---|---|---|
| Low | <1GHz | Coverage | Large | Net Coverage, IoT | Long range coverage, cost effective infrastructure |
| Mid | 1-6 GHz | Macro Capacity | Small | Net coverage, capacity for data transfer | Short range compared to higher frequency, reduced performance |
| High | >6 GHz | Hotspot Capacity | Ultra small | Capacity for very high data transfer | Short range, low latency, allows high speed data transfer |

The range of FR1 is from 450MHz to 6GHz. The frequencies of less than 3GHz are called sub-3GHz, and others are called C-Band. The duplex modes used in FR1 are TDD and FDD. The FR2 (mm-wave) range is from 24.25GHz to 52.6 GHz. It supports three spectrum types: licensed, unlicensed, and shared, and designed to connect virtually, including objects, machines, and devices. It provides far larger network capacity, higher data speeds, virtual reality, augmented reality, increased availability, more reliability, ultra-low latency since it combines traditional RF bands and new radio bands. The FR2 is using TDD duplex mode.

## IV. Structure of the Skin

The skin is the largest organ of the body. Three layers make up the skin, including the epidermis, dermo, and fat layers. Specific tasks are carried out by each layer. The epidermis is the thickest, outermost layer of skin. The epidermis' topmost layer, the stratum corneum, is water-resistant and, while intact, keeps maximum unfamiliar organisms outside the body. Melanocytes are cells that are dispersed throughout the epidermis's basal layer and produce the pigment melanin, which is a major factor in skin color. However, the main job of melanin is to block out ultraviolet sunlight, which harms DNA and has several negative effects, such as skin cancer.

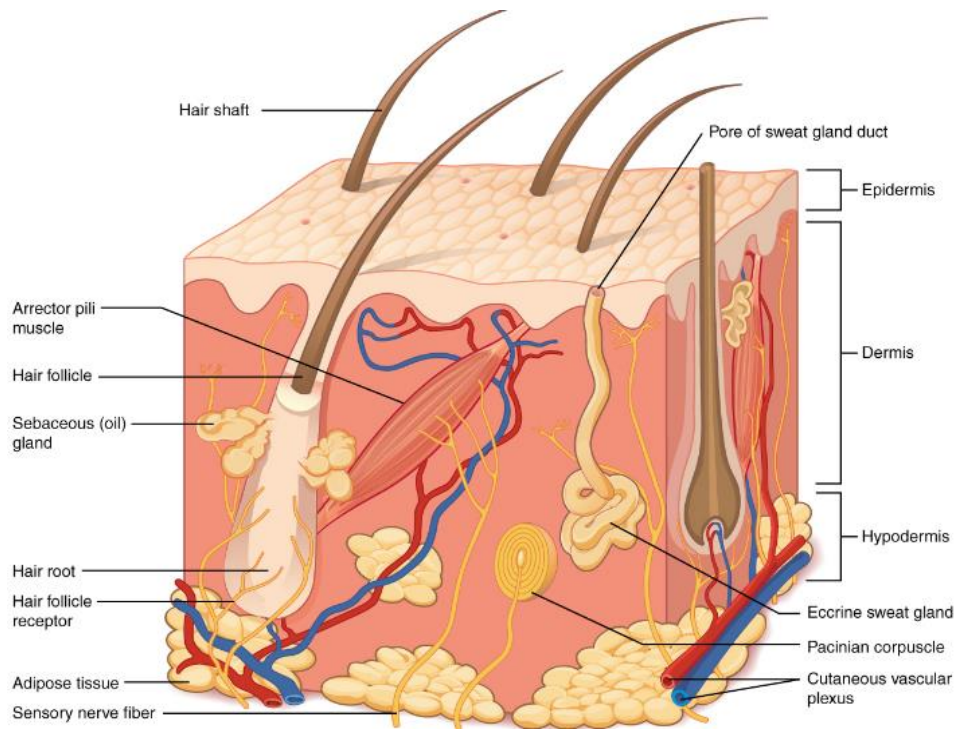

*Figure 2 **Layers of the Skin** (OpenStax, n.d.)*

The second layer is dermis contains fibrous tissue which provides skin elasticity. The dermis is composed to blood vessels, hair follicles, sweat glands, and nerve endings. A layer of fat that acts as an energy storage area, protective padding, and heat and cold insulation for the body is located below the dermis. Living cells known as fat cells, which are connected by fibrous tissue, contain the fat. Figure 2 **Layers of the Skin** shows the layers of the skin.

Function of the Skin

The skin is a rich source of several hormone types and immune-active compounds in addition to being the target organ of the neurological system and neurohormonal system. The skin can be thought of as a peripheral, semi-autonomous "organ" that guards the remainder of the body against harm from mechanical forces, dehydration, heat, and cold. Additionally, a variety of receptors and sensors are built into the skin to monitor the environment for important environmental parameters.

For upstream and downstream communication with the central nervous system, the skin possesses a robust peripheral nerve network. Additionally, there is communication through the circulation and immune system signaling molecules. The protection provided by the skin is among its most significant roles. The first line of defense for our body is therefore thought to be our skin. The epidermis, or outermost layer of skin, is responsible for this resistance. The health of the epidermis is essential since it defends against bacteria, viruses, and other dangerous organisms. Additional functions of skin are temperature regulation, moisture retention, toxin removal and vitamin D production.

Penetration of EMF on human Skin

With much too high levels, electromagnetic radiation and fields are becoming denser every day. These reactions—known as EHS symptoms. No living thing is immune to EMF. Most people are not aware of EMF sensitivities, thus neither they nor their general practitioners can identify the symptoms. Cancer, diabetes, and heart disease are later stages of symptoms and responses. Some folks may be able to get away from it or begin using EMF protection. Nobody or nothing can shield themselves from or avoid 5G. There is no cream that can protect the skin from 5G because every cream also contains water, which is an EMF transmitter. The cream is a paint and is unsafe for various reasons if it totally covers the skin. The penetration depth of 5G on human skin is greater than that of earlier generations. A full coat of body paint can be damaging to your health if you are not careful, despite being a usually harmless hobby. Full-body paint can cause heatstroke because it closes your pores and interferes with your body's natural cooling and sweating processes. Cancer is one of the anticipated health implications of 5G on skin. EMF can penetrate the human body by two ways, thermal and non-thermal.

a) Thermal Effects

In medicine, RF is used for therapeutic purposes; indeed, some frequencies that exist thermal effects. It includes wound healing, cerebral ischemia, nerve regeneration, graft behavior, myocardial and other conditions. Some common health issues caused by thermal effects are blindness, heating, sterility, and a burning sensation. Stretched skin tissue serves as a receiver for millimeter radiation. Blood veins, intercellular fluid-filled cavities, and particularly sweat gland drainage ducts can all be examples of these. Such a duct's helicoidally formed end, which is packed with conductive sweat moisture, is ideal for use as an MMW antenna. It is possible to anticipate that they may overheat, harming both the cell and the tissues around it. Because organic structures are not a homogeneous conductive mass, it is considerably worse. The outer millimeter of the brain mass was warmed to 35°C in a model experiment using dead bovine brains after receiving a 30-minute dose of a 39 GHz field. That results in warmth that is roughly 175 times stronger than radiation with a 1.9 GHz field. This is due to the micro-architecture of brain tissue. However, this demonstrates how the ICNIRP concepts have been oversimplified.

Warming is viewed as a straightforward process of heat dispersion; heat diffusion varies in intricate composite fabric layers.

b) Non-Thermal Effects

The detrimental effects of RF-EMF's non-thermal biological interactions with both people and animals are not mentioned in the ICNIRP guidelines. Issues caused by non-thermal effects are continuous pulse modification, communication of electrical and chemical signal changes, and carcinogenic possibilities. During the transmission of EMF from one device to another, it can be replicated, fascinated, and diverted, depending on the source frequency and the uncovered body penetration. It deals with the modification of biological processes whose functions are not affected by temperature and the physical impacts on the structure of biomolecules. Non-thermal effects of EMF are characterized by the fact that they happen at field levels that are much lower than the ICNIRP exposure limits.

Impact of EMF on skin

The EMR is generated from many sources like blue light, UV, and IR radiation, significantly affecting the skin [9]. The 28 GHz frequency from 5G EMR can be able to penetrate our body up to 2 mm. The primary target is damaging the skin [10]. The skin produces hyperpigmentation by inducing melanogenesis, which is prompted by UV rays and the blue light of EMR [11]. 5G EMR can generate ROS, which causes skin damage, rough texture, hyperpigmentation, and aging wrinkles [12]. However, a few factors, including air pollution and UV radiation, can cause the creation of ROS in the skin. The temperature of the human body gradually increases because of RF-EMF. The range of the outer surface of human skin is between 30°C and 35°C. It can damage the protein induction and cells of the skin. When the temperature reaches 43°C, the pain will be identified, affecting the eyes, peripheral nerves, sweat glands, and testes [14].

EMF Exposure Concerns

Two changes will probably occur concerning exposure to the human body. 1. BS will operate by a massive number of transmitters. 2. Using narrower beams as a solution at high frequency for higher attenuation. 5G allows a higher mobile data volume per geographical area for an increasing number of devices and needs more infrastructure. The shorter ranges 24-100GHz, require higher frequencies. IARC defined the possibly carcinogenic to humans and animals in the frequency range from 30KHz to 300GHz of RF-EMF. In terms of health, the development of 5G will result in a rise in cancer cases, particularly in the brain. [18]. Several kinds of research are ongoing about the risk of cancer using cellular equipment and EMF generation. Due to insufficient information, we cannot conclude that a radio frequency emission causes no health issues.

## V. EMF Exposure Standards

Some projects on pre-generation EMF emissions are i) LEXNET which provides the guidelines on configurations of antennas; ii) MONICEM, which monitors and controls the Base Stations EMF generation. But these guidelines are only supported in Italy [16] and exist two different limits i) general limits and ii) restrictive limits (school, house, etc.) when prolonged and continuous-time usage.

ICNIRP

Reference levels for continuous whole-body exposures were provided by ICNIRP in 1998. It does not include all restrictions of every kind. In 2020, it offered RF-EMF strategies for preventive exposure to EMF to protect the public from RF-EMF between 100 KHz and 300 GHz. It covers many applications like BS, WiFi, mobile devices, and Bluetooth in 5G technologies. Many countries accept the conventional limits of EMF by ICNIRP, and some countries like Russia, China, Canada, India, and Poland implement national laws for strict limits. The ICNIRP does not provide recommendations for EMF exposure in terms of SAR at frequencies above 10GHz. It produces the consequences of exposure limitations with a vast boundary of protection. Two types of restrictions are basically used while limiting exposure 1) Reference levels correspond to fields in the environment, which can be measured simply, while 2) basic constraints refer to fields in humans, which cannot be quantified readily. And it is frequently applied to guarantee safety. According to scientific basis, the health effects of skin stimulation up to ~10 MHz increase the body's temperature from ~100 KHz. When the body temperature increase $>1°C$ and the tissue temperature increase $> 41°C$, the health effects are identified.

FCC

The United States Federal Communications Commission suggests that devices operating higher than 6GHz frequencies use Power Density as a metric to evaluate EMF. But it is not applicable when the devices work close to the body [19]. The FCC does not have any regulations regarding SAR exposure to EMF at frequencies above 6GHz. The Federal

Communications Commission proposes Power Density as a measurement system to calculate the person's exposure to Electro Magnetic fields produced by transmitting devices above 6 GHz. Still, PD is inefficient when the devices are nearby to the person's body [8]. The FCC procedures include Maximum Permissible Exposure limits in order of EMFPD for individuals functioning sources at a 300 kHz and 100 GHz frequency range [14].

EMF Exposure Limits

The recommendations made by WHO and ITI were based on guidelines supplied by ICNIRP that were generated from observations of the thermal effects that EMFs have on the human body. The impacts that can be seen include heating and induced current (<300 MHz), heating of the body (300 MHz to 10GHz), and heating of the skin (>10GHz). 5G operates less than 1GHz to coverage in suburban and rural areas, from 1GHz to 6GHz offers both capacity and coverage, and above 6GHz provides very high data rates. Based on the danger of glioma and a malignant form of brain cancer, the World Health Organization has classed RF-EMF fields as having potential human carcinogenicity. In any case, people should not experience any detrimental health effects if the EMF fields are employed within the parameters of ICNIRP.

## VI. EMF Evaluation Metrics

To assess exposure to 5G, we need to concern about the continuous changes in the activity of UE and BS related to MIMO technology. Koprivica et al. show that the EMF levels differ depending on time and space using GSM downlink band 900 MHz, and they measured the EMF levels in Serbia by using GSM and UMTS and showed that it exceeds the limits in15.6% of the locations [15]. In some regions, the level of EMF emission is below the limits. To estimate the level of EMF, we need to consider some factors such as scheduling time, utilization of Base Station, time-division duplex, and user's spatial distribution.

Several protocols are available to measure the EMF rate during uploading and downloading, and it guarantees a safe level of EMF exposure. Depending on the frequency and length of exposure, different metrics are employed to measure temperature. For local directions, for instance, the power density is used at higher frequencies and the absorbed energy rate (SAR) at lower frequencies. In comparison to Power Density (PD), the Specific Absorption Rate (SAR) is a more useful statistic for assessing effects on human health. Human tissues are penetrated by electromagnetic waves, which have the power to influence molecular energy levels. The quantity of energy that the human body perceives from EMF can be shown by SAR.

Compared to SAR, PD cannot evaluate the reflected effect adequately. SAR is the more suitable metric to evaluate human skin temperature [20].

i) Power Density

Two methods are used to estimate power density generally. i) E and H field based ii) plane wave equivalent approximation based. The surface and poynting vectors are evaluated when the PD is based on E and H field. The poynting vector is given by

$$S = E \times H^* \qquad (1)$$

This method is mainly used for near-field calculation. It is appropriate to use the plane wave equivalent approximation to determine far-field power density. In this supposition, the E and H fields are connected to a scalar constant: $H = \frac{\hat{r} \times E}{\eta}$ where $\hat{r}$ is the radial direction vector, and $\eta$ is the free space wave impedance.

To calculate PD using power radiated at a distance $d$ by [21],

$$PD(d) = \frac{|E(d)|^2}{\rho_0} \; [W / m^2] \qquad (2)$$

Where $\rho_0$ denotes free space impedance and $E(d)$ denotes complex amplitude of the incident electric field. By using transmitter's constraints, the above will be rewritten as

$$PD(d, \phi) = \frac{P_T G_T(d, \phi)}{4\pi d^2} \qquad (3)$$

Where $P_T$ represents transmit power, $G_T$ for transmit antenna gain, and $d$ stands for the BS-UE distance in meters.

ii) Specific Absorption Rate

At the level of higher frequencies such as 28GHz the emissions from RF will be deposited to the thin surface area of the human skin. It is defined as the absorption energy rate of human body from an EMF by the unit of mass and time. The formula for the local SAR value at a point $p$ is

$$SAR(p) = \frac{\sigma |E(p)|^2}{\rho} \; [W/kg] \qquad (4)$$

Where material density is denoted as $\rho$ and $\sigma$ is defined as the material conductivity. To calculate SAR at the point of air-skin boundary using the following formula

$$SAR(d,\phi) = \frac{2PD(d,\phi)(1-R^2)}{\delta\rho} \quad (5)$$

Where $\delta$ is denoted as the penetration depth of the skin (used $10^{-3}$ m), $\rho$ is denoted as the mass density of the tissue (used 1g/cm$^3$) and $R$ is denoted as the coefficient of the reflection [22]. The variable $d$ and $\phi$ depends upon the user equipment position in a cell. Depending on the type of tissue considered, SAR's value will change. For instance, a tissue's SAR value for the eyes will be different than a tissue's SAR value for the limbs. Additionally, the exposed tissue has a distinct value for SAR and SAR deep.

Reduction of human EMF Exposure

The level of EMF exposure can be decreased by coordinating and using multiple spectrum bands. Because of this, a wireless system should have a lower cell size with a higher carrier frequency, posing more risks to human health. The human exposure can be decreased when the BS implements a power control or an adaptive beamforming technology.

## VII. Result Analysis

Table 2 compares the EMF penetration depth into the human skin of three wireless systems—3.9G, 4G, and 5G. The epidermis and dermal layers of the skin absorb more than 90% of the electromagnetic energy that is transmitted, showing that the skin may be the primary target of 5G radiation. The depth of EMFs' penetration into the skin gradually dropped as frequency increased.

*Table 2: EMF Penetration Level on Skin-Air Boundary*

| Network Type | SAR (W/Kg) | Penetration Depth (mm) |
|---|---|---|
| 5G | 1.24 | 0.9 |
| 4 G | 0.000644 | 21 |
| 3.9 G | 0.0000435 | 23 |

SAR levels vary depending on several key factors, including frequency and material type. This graph displays the skin's SAR value at 10 cm from a transmitting device to an uplink. Since high frequency EMF cannot penetrate deeply into human skin, it does not mean that it is not dangerous to the human.

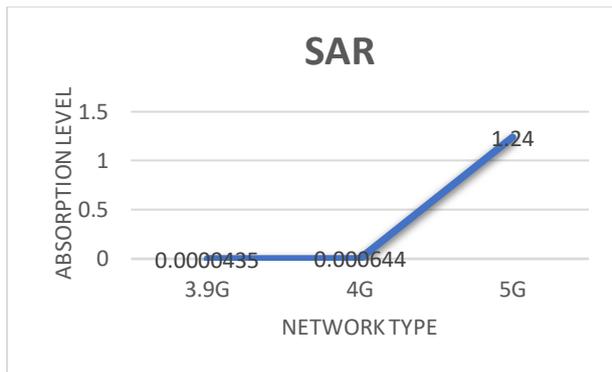

*Figure 3 Comparison of skin absorption level of EMF*

Using the Specific Absorption Rate (SAR) calculation at the air-skin boundary, Figure 3 compares the absorption level of EMF on human skin. It is calculated by using SAR formula. Comparing to 3.9 G and 4G, the level of penetration is high in 5G.

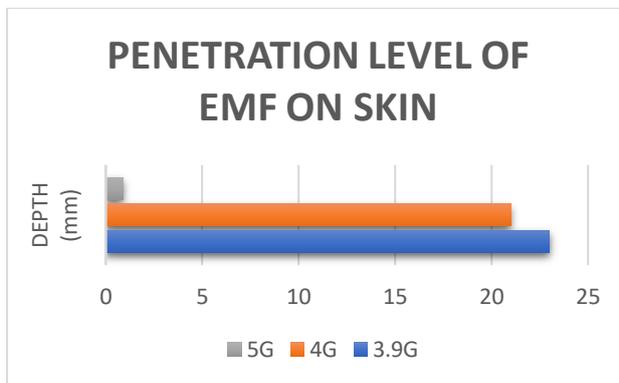

*Figure 4 : Penetration Depth on Skin*

Figure 4 produce the penetration level of EMF using SAR and depth is measured in mm. Within the concentrated area, the SAR may be higher, which could lead to subsequent health issues including skin heating.

**Conclusion**

With its higher radiation levels, the 5G network is now supporting a large range of services on demand and increasing safety questions for humans and other living things. Concerns regarding the negative effects of 5G on the environment and human health were expressed by many researchers. EMF radiations and exposure levels will be raised due to a growing number of access points for 5G as well as an increase in the number of mobile devices. There is still a debate in science on adverse health effects associated with electromagnetic fields. In the field of 5G safety guidelines have been developed by various health organizations, including a minimal distance between User Equipment and Base Station or transmitters. It is dangerous for people when used at a high frequency. In comparison to networks from earlier generations, 5G features high frequencies and a low penetration. But if the frequencies are increased from the sub-band of 6GHz, it will impact on the skin layers. This paper analyses the penetration level of 5G EMF on human skin by measuring Specific Absorption Rate and the exposure of EMFs is investigated especially skin diseases. It analyses the impact of EMF penetration in human skin layers and discusses how to reduce exposure levels for 5G.

References:

[1]. Kyuri Kim, Young Seung Lee, Nam Kim, Hyung-Do Choi, Kyung-Min Lim, "5G Electromagnetic Radiation Attenuates Skin Melanogenesis In Vitro by suppressing ROS Generation" Antioxidants (Basel), Aug 2022, 11(8):1449.

[2]. Kyuri Kim, Young Seung Lee, Name Kim, Hyung-Do Choi, Dong-Jun Kang, Hak Rim Kim, Kyung-Min Lim, "Effects of Electromagnetic Waves with LTE and 5G Bandwidth on the Skin Pigmentation In Vitro", International Journal of Molecular Sciences, Dec 2020, 22(1):170.

[3]. Mehdizadeh AR, Mortazavi SMJ, "5G Technology: why should we expect a shoft from RF-induced brain cancers to skin cancers?", Journal of biomedical physics engineering, 2019, 9(5), 505-506.

[4]. Halliwell B. Oxidative stress and cancer: have we moved forward? Biochem J. 2007;401:1–11. doi: 10.1042/BJ20061131.

[5]. Mortazavi S A R, Mortazavi G, Mortazavi S M J. Use of cell phones and brain tumors: a true association? Neurol Sci. 2017;38:2059–60. doi: 10.1007/s10072-017-3055-x.

[6]. Mortazavi S, Mortazavi S, Haghani M. Evaluation of the validity of a Nonlinear J-shaped dose-response relationship in cancers induced by exposure to radiofrequency electromagnetic fields. Journal of Biomedical Physics and Engineering. 2019;9 doi: 10.31661/jbpe.v0i0.771.

[7]. Karipidis K., Mate R., Urban D., Tinker R., Wood A. 5G mobile networks and health—A state-of-the-science review of the research into low-level RF fields above 6 GHz. J. Expo. Sci. Environ. Epidemiol. 2021;31:585–605. doi: 10.1038/s41370-021-00297-6. –

[8]. Betzalel N., Ishai P.B., Feldman Y. The human skin as a sub-THz receiver–Does 5G pose a danger to it or not? Environ. Res. 2018;163:208–216. doi: 10.1016/j.envres.2018.01.032.

[9]. Meinke M.C., Busch L., Lohan S.B. Wavelength, dose, skin type and skin model related radical formation in skin. Biophys. Rev. 2021;13:1091–1100. doi: 10.1007/s12551-021-00863-0.

[10]. Sutterby E., Thurgood P., Baratchi S., Khoshmanesh K., Pirogova E. Evaluation of in vitro human skin models for studying effects of external stressors and stimuli and developing treatment modalities. View. 2022;3:20210012. doi: 10.1002/VIW.20210012.

[11]. Campiche R., Curpen S.J., Lutchmanen-Kolanthan V., Gougeon S., Cherel M., Laurent G., Gempeler M., Schütz R. Pigmentation effects of blue light irradiation on skin and how to protect against them. Int. J. Cosmet. Sci. 2020;42:399–406. doi: 10.1111/ics.12637.

[12]. Liebel F., Kaur S., Ruvolo E., Kollias N., Southall M.D. Irradiation of skin with visible light induces reactive oxygen species and matrix-degrading enzymes. J. Investig. Dermatol. 2012;132:1901–1907. doi: 10.1038/jid.2011.476.

[13] Rik Roelandts "Cellular Phones and the skin", Dermatology, 2003,207:3-5.

[14] Seungmo Kim and Imtiaz Nasim, "Human Electromagnetic Field Exposure in 5G at 28GHz", IEEE Consumer Technology Society, 2162-2248, May 2020, PG-41.

[15] M. Koprivica, M. Petric, M. Popoviĺc, J. Milinkoviĺc, S. Nikšiĺc, and ´A. Neškovic, "Long-term variability of electromagnetic field strength for ´ gsm 900mhz downlink band in belgrade urban area," in Proc. of 22nd Telecommunications Forum Telfor (TELFOR), Belgrade, Serbia, pp. 9– 12, IEEE, 2014.

[16] Luca Chiaraviglio, Angela Sara Cacciapuoti, Gerardo Di Martino, Marco Fiore, Mauro Montesano, Damiano Trucchi, Nicola Blefari-Melazzi, "Planning 5G Networks under EMF Constraints: State of the Art and Vision", IEEE Access, 2169-3536, 2018.

[17] D.Plets, W.Joseph, K.Vanhecke,, L.Martens, "Exposure Optimization in indoor wireless networks by heuristic network planning", Progress in Electromagnetic Research, vol.139, pp.445-478, 2013.

[18] A. J. Swerdlow, M. Feychting, A. C. Green, L. Kheifets, D. A. Savitz, and International Commission for Non-Ionizing Radiation Protection Standing Committee on Epidemiology, "Mobile phones, brain tumors, and the interphone study: where are we now?," Environmental health perspectives, vol. 119, no. 11, p. 1534, 2011.

[19] T.Wu, T.Rappaport and C.Collins, "Safe for generations to come: Considerations of safety for millimetre waves in wireless waves in wireless communication", IEEE Microwave Magazine, vol.16, no.2, 2015.

[20] Imtiaz Nasim and Seungmo Kim, "Mitigation of Human EMF Exposure in a cellular Wireless System" IEEE Vehicular Technology Conference, 2020.

[21] Umar Hasni, ErdemTopsakal, Wearable Antennas for On-Body Motion Detection, URSI 2020, 978-1-7281-6670-4/20/$31.00 ©2020 IEEE.

[22] T.Wu, T.Rappaport and C.Collins, "The human body and millimetre wave wireless communication systems: interactions and implications" in Proc. IEEE International Conference on Communications, 2015.